\documentclass[a4paper,11pt]{article}
\usepackage{jheppub} % for details on the use of the package, please
\usepackage[T1]{fontenc} % if needed

\usepackage{mathtools,mathstyle,breqn}
\usepackage{amsmath}
\usepackage{amssymb}
\usepackage[utf8]{inputenc}
\usepackage{graphicx}
\usepackage{subfig}
\usepackage{blindtext}

\title{\boldmath Electroweak Phase Transition, Gravitational Waves and Dark Matter in Two Scalar Singlet Extension of The Standard Model}

\author[a]{Vahid Reza Shajiee}
\author[a]{Ali Tofighi}

\affiliation[a]{Department of Physics, Faculty of Basic Sciences, University of Mazandaran,\\Babolsar, Iran}
\emailAdd{v.shajiee@stu.umz.ac.ir}
\emailAdd{a.tofighi@umz.ac.ir}


\abstract{In this paper, the electroweak phase transition, the gravitational waves and the dark matter issues are investigated in two scalar singlet extension of the standard model. The detectability of the gravitational wave signals are discussed by comparing the results with the sensitivity curves of $\mathbf{eLISA}$, $\mathbf{ALIA}$, $\mathbf{DECIGO}$ and $\mathbf{BBO}$ detectors. It is shown that the results support the recent reports on the dark matter relic density by $\mathbf{Planck}$ $\mathbf{2018}$ collaboration and the direct detection experiment by $\mathbf{XENON1T}$ $\mathbf{2018}$ collaboration.}

\begin{document}
\maketitle
\flushbottom

\section{Introduction}
\label{sec:1}

The failure of the Standard Model (SM), in describing phenomena like the baryon asymmetry of the universe (BAU) and the dark matter (DM), brings to mind that the SM cannot be considered as a fundamental model. Nevertheless, the discovery of the Higgs boson~\cite{Aad:2012tfa,Chatrchyan:2012xdj} as the first observed scalar has opened the way to consider the SM as an effective field theory (EFT) and also a window to the Higgs portal. To address the BAU and the DM problems, many theories and models have been proposed beyond the SM such as supersymmetry studies~\cite{Cline:2000kb,Huber:2006wf,Huber:2006ma,Pietroni:1992in,Davies:1996qn,Ham:2004nv,Menon:2009mz,Carena:2012np,Huber:2007vva,Kozaczuk:2014kva,Kozaczuk:2012xv,Jungman:1995df,Menon:2004wv,Cirigliano:2006dg,Cao:2011re}. Due to the attraction of the Higgs portal, it has been always of interest to investigate the SM extensions which directly challenge the Higgs portal like multi-scalar extensions~\cite{Damgaard:2015con,Vaskonen:2016yiu,Beniwal:2017eik,Chen:2017qcz,Espinosa:2011ax,Cline:1996mga,Fromme:2006cm,Kang:2017mkl,Dorsch:2013wja,Haarr:2016qzq,Gunion:1989ci,FileviezPerez:2008bj,Alanne:2014bra}. The existence of interactions between the Higgs and new scalars makes such models reasonable for explaining the BAU, which needs a strong first-order electroweak phase transition (SFOEWPT), the gravitational waves (GW) produced by an SFOEWPT and the DM. Moreover, such models also have other benefits. First, they are simple and straightforward. Second, they may be renormalized, so no new physics scale is needed. Third, they may be gauge independent, if there exists a barrier in the potential at tree-level~\cite{Patel:2011th}.

To justify the BAU, there is a need for Baryogenesis to exist~\cite{Kuzmin:1985mm,Shaposhnikov:1987tw,Dine:2003ax,Cline:2006ts,Canetti:2012zc} which itself needs an SFOEWPT, i.e. $\frac{v_{c}}{T_{c}}\gtrsim1$ where $v_{c}$ is the Higgs vacuum expectation value (VeV) at critical temperature $T_{c}$. This would not happen in the SM, but adding one or more new scalars to the SM potential may lead to an SFOEWPT. With regard to the new potential structure, two different phase transitions (PT) can happen. One of them is one-step PT in which there only exist initial and final phases. Cooling down the universe, it goes through a phase transition and breaks the electroweak symmetry. The other one is two-step (or multi-step) PT in which there also exists an intermediate phase (or more) between initial and final phases~\cite{Land:1992sm,Hammerschmitt:1994fn,Patel:2012pi,Huang:2014ifa,Blinov:2015sna}. The reader is referred to~\cite{Ellis:2018mja,Baker:2017zwx,Croon:2018erz,Beniwal:2018hyi,Huang:2017laj,Hashino:2018zsi,Mazumdar:2018dfl,Ghosh:2017fmr} for the most recent studies on the EWPT.

During the SFOEWPT, the bubbles with the non-zero VeV nucleates in the plasma. The stochastic GW background arising from the SFOEWPT can be generated by the bubbles collisions and shocks~\cite{Kosowsky:1991ua,Kosowsky:1992rz,Kosowsky:1992vn,Kamionkowski:1993fg,Caprini:2007xq,Huber:2008hg}, the sound waves~\cite{Hindmarsh:2013xza,Giblin:2013kea,Giblin:2014qia,Hindmarsh:2015qta}, and the Magnetohydrodynamic (MHD) turbulence~\cite{Caprini:2006jb,Kahniashvili:2008pf,Kahniashvili:2008pe,Kahniashvili:2009mf,Caprini:2009yp} in the plasma. Since the EWPT in the SM is a cross-over instead of strong one, the SM cannot predict the GW produced by the EWPT. So, this is another reason to look for beyond the SM. The recent observations of astrophysical GW~\cite{Abbott:2016blz,Abbott:2016nmj,Abbott:2017ylp,Abbott:2017vtc,Abbott:2017oio,TheLIGOScientific:2017qsa,Monitor:2017mdv,Abbott:2017gyy} have brought the hope to detect the GW produced by the EWPT~\cite{Caprini:2015zlo,Weir:2017wfa,Caprini:2018mtu}. The reader is referred to~\cite{Ellis:2018mja,Croon:2018erz,Beniwal:2018hyi,Huang:2017laj,Hashino:2018zsi,Demidov:2017lzf,Mazumdar:2018dfl,Kobakhidze:2016mch,Kobakhidze:2017mru,Dev:2016feu} for the most recent studies on the GW produced by the EWPT.

As mentioned before, the SM cannot explain DM which existence is well established by cosmological evidence. As the simplest way, this incompetence can be justified by adding one (or more) gauge singlet scalar to the SM. Since the DM should be stable to provide the observed relic density $\Omega_{c}h^{2}=0.120\pm0.001$ by $\mathbf{Planck}$ $\mathbf{2018}$~\cite{Aghanim:2018eyx}, it is necessary to impose a discrete symmetry on the DM candidate, in present study $S_{2} \rightarrow -S_{2}$. On the other hand, the global minimum of potential at zero temperature spontaneously breaks this discrete symmetry, so necessarily $<S_{2}>=0$. The reader is referred to~\cite{Athron:2018ipf,Bernal:2018ins,Baum:2017enm,Baker:2017zwx,Croon:2018erz,Beniwal:2018hyi,Li:2018qip,Hashino:2018zsi,Bandyopadhyay:2017tlq,Yepes:2018zkk,Ghosh:2017fmr} for the most recent studies on the DM.

The present work is arranged as follows: In section \ref{sec:2}, the most general and renormalizable extension of the SM is presented by adding two scalar sectors $S_{1}$ and $S_{2}$ to the usual SM potential\footnote{The model first presented in~\cite{Tofighi:2015fia} without the GW and the DM discussions. Here, the results of~\cite{Tofighi:2015fia} are improved for the EWPT, also, the GW and the DM signals are investigated.}. Assigning a non-zero VeV to $S_{1}$, the SFOEWPTH can occur in the model. Imposing a $Z_{2}$ symmetry on $S_{2}$ makes it a viable candidate for the DM. Also, constraints on the parameter space are discussed. The EWPT, GW and DM are respectively investigated in sections \ref{sec:3}, \ref{sec:4} and \ref{sec:5}. Finally, some conclusions are presented in section \ref{sec:6}.

\section{The Model}
\label{sec:2}

The tree-level potential of the model is given by
\begin{equation} \label{eq:2.1}
\begin{split}
  V = & - m^{2} H^{\dagger}H + \lambda (H^{\dagger}H)^2 + \kappa_{0} S_{1} + 2 (\kappa_{1} S_{1}+\kappa_{2} S_{1}^{2}+\kappa_{3} S_{2}^{2}) H^{\dagger}H \\
  & + \frac{1}{2} m_{1}^{2} S_{1}^{2} + \frac{\lambda_{1}}{4} S_{1}^{4} + \kappa_{4} S_{1}^{3} + \frac{1}{2} m_{2}^{2} S_{2}^{2} + \frac{\lambda_{2}}{4} S_{2}^{4} + \kappa_{5} S_{1} S_{2}^{2},
\end{split}
\end{equation}
The potential \ref{eq:2.1} is the usual SM potential with two extra gauge singlet scalars and interaction terms which provide Higgs portal between the new scalars and the usual SM particles. H stands for the complex Higgs doublet, $H=\begin{pmatrix}\chi_{1}+i\chi_{2} \\ \frac{1}{\sqrt{2}}(h+i\chi_{3}) \\ \end{pmatrix}$. $S_{2}$ stands for the DM imposing $S_{2}\rightarrow-S_{2}$. Acquiring a non-zero VeV, $S_{1}$ improves the strength of EWPT. The linear term of $S_{1}$ can be neglected by a shift in the potential. The $Z_{2}$ symmetry forbids the existence of linear and cubic terms for $S_{2}$, so the equation \ref{eq:2.1} is the most general renormalizable potential which could be made by adding two new scalars. In the unitary gauge at zero temperature, the theoretical fields can be reparameterized in terms of the physical fields,
\begin{equation} \label{eq:2.2}
H=\begin{pmatrix}0 \\ \frac{1}{\sqrt{2}}(h+v) \\ \end{pmatrix},\quad S_{1}=s_{1}+\chi,\quad S_{2}=s_{2},
\end{equation}
where $v=246.22 (GeV)$ and $\chi$ are the Higgs and $S_{1}$ VeV, respectively. Without loss of generality, one can write
\begin{equation} \label{eq:2.3}
\begin{split}
V = & -\frac{1}{2} m^{2} h^{2} + \frac{\lambda}{4} h^{4} + (\kappa_{1} s_{1}+\kappa_{2} s_{1}^{2}+\kappa_{3} s_{2}^{2}) h^{2} \\
  & + \frac{1}{2} m_{1}^{2} s_{1}^{2} + \frac{\lambda_{1}}{4} s_{1}^{4} + \kappa_{4} s_{1}^{3} + \frac{1}{2} m_{2}^{2} s_{2}^{2} + \frac{\lambda_{2}}{4} s_{2}^{4} + \kappa_{5} s_{1} s_{2}^{2}.
\end{split}
\end{equation}
In order to have a stable potential, it is required that~\cite{Espinosa:2011ax,Tofighi:2015fia}
\begin{equation} \label{eq:2.4}
\lambda>0,\quad \lambda_{1}>0,\quad \lambda_{2}>0,\quad \kappa_{2}>-\frac{\sqrt{\lambda \lambda_{1}}}{2},\quad \kappa_{3}>-\frac{\sqrt{\lambda \lambda_{2}}}{2}.
\end{equation}
The tadpole equations at $(v,\chi,0)$ read
\begin{equation} \label{eq:2.5}
\begin{split}
& m^{2}=\lambda v^{2} + 2 (\kappa_{1} \chi + \kappa_{2} \chi^{2}), \\
& m_{1}^{2}=-\lambda_{1} \chi^{2} - 3 \kappa_{4} \chi - 2 \kappa_{2} v^{2} - \frac{\kappa_{1} v^{2}}{\chi}.
\end{split}
\end{equation}
From the diagonalization of squared-mass matrix and the tadpole equations, one can get
\begin{equation} \label{eq:2.6}
\begin{split}
& \lambda=\frac{M_{1}^{2}sin^{2}(\theta)+M_{H}^{2}cos^{2}(\theta)}{2 v^{2}}, \\
& \kappa_{2}=\frac{(M_{H}^{2}-M_{1}^{2})sin(2\theta)}{8v\chi}-\frac{\kappa_{1}}{2\chi}, \\
& \lambda_{1}=\frac{1}{2\chi^{2}}\left(M_{1}^{2}cos^{2}(\theta)+M_{H}^{2}sin^{2}(\theta)+\frac{\kappa_{1}v^{2}}{\chi}-3 \chi \kappa_{4}\right), \\
& m_{2}^{2}=M_{2}^{2}-2\kappa_{3}v^{2}-2\kappa_{5}\chi,
\end{split}
\end{equation}
where $M_{H}=126 (GeV)$, $M_{1}$, $M_{2}$ and $\theta$ are the Higgs mass\footnote{The last announcement for the Higgs mass is $M_{H}=125.09 (GeV)$~\cite{Aad:2015zhl}, however, 1-3 GeV deviation is acceptable.}, the physical mass of $S_{1}$, the physical mass of $S_{2}$ (the DM mass) and the mixing angle, respectively. In Ref.~\cite{Profumo:2014opa}, by performing a global fit to the Higgs data from both $\mathbf{ATLAS}$ and $\mathbf{CMS}$, the constraint on the mixing angle was given $|\theta|\leq32.86^{\circ}$ at $95\%$ confidence level (CL). In Ref.~\cite{Chao:2016vfq}, by performing a universal Higgs fit, the upper limit on the mixing angle was given $|\theta|\leq30.14^{\circ}$ at $95\%$ CL. In the present work, a Monte Carlo scan is performed over the parameter space with
\begin{equation} \label{eq:2.7}
\begin{split}
& 5GeV \leq M_{1} \leq 750 GeV,\quad 5GeV \leq M_{2} \leq 750 GeV,\quad -23^{\circ}\leq \theta \leq23^{\circ}, \\
& -80GeV\leq \kappa_{1}\leq 80GeV,\quad 0.0001\leq \kappa_{3}\leq 0.1,\quad -80GeV\leq \kappa_{4}\leq 80GeV, \\
& -80GeV\leq \kappa_{5}\leq 80GeV,\quad 30 GeV\leq \chi\leq 120 GeV,\quad 0\leq \lambda_{2}\leq 4.
\end{split}
\end{equation}

\section{Electroweak Phase Transition}
\label{sec:3}

To investigate the EWPT in a model, one needs the finite temperature effective potential given by
\begin{equation} \label{eq:3.1}
V_{eff}=V_{tree-level}+V_{1-loop}^{T=0}+V_{1-loop}^{T\neq0},
\end{equation}
where $V_{tree-level}$, $V_{1-loop}^{T=0}$ and $V_{1-loop}^{T\neq0}$ are the tree-level potential \eqref{eq:2.3}, the one-loop corrected potential at zero temperature (the so-called Coleman-Weinberg potential) and the one-loop finite temperature corrections, respectively. The last two read
\begin{equation} \label{eq:3.2}
\begin{split}
& V_{1-loop}^{T=0}=\pm\frac{1}{64\pi^2}\sum_{i=h,s_{1},s_{2},W,Z,t} n_i
m_i^4\left[\log\frac{m_i^2}{Q^2}-C_i\right],\\
& V_{1-loop}^{T\neq0}=\frac{T^4}{2\pi^2} \sum_{i=h,s_{1},s_{2},W,Z,t} n_i J_\pm \left[ \frac{m_{i}^{2}}{T^{2}}\right],\\
& J_\pm(\frac{m_{i}^{2}}{T^{2}}) = \pm \int_0^\infty dy\, y^2 \log\left(1\mp e^{-\sqrt{y^2+\frac{m_{i}^{2}}{T^{2}}}}\right),
\end{split}
\end{equation}
where $n_{i}$, $m_{i}$, $Q$ and $C_{i}$ denote the degrees of freedom, the field-dependent masses, the renormalization scale and the numerical constants, respectively. The degrees of freedom and the numerical constants are respectively given by $(n_{h,s_{1},s_{2}},n_{W},n_{Z},n_{t})=(1,6,3,12)$ and $(C_{W,Z},C_{h,s_{1},s_{2},t})=(5/6,3/2)$. The upper (lower) sign is for bosons (fermions). Assuming the longitudinal gauge bosons polarizations are screened by plasma, thermal masses just contribute to the scalars, so Daisy corrections become small and can be neglected. There are three possibilities to deal with the renormalization scale $Q$. First one is to add some counter terms to the effective potential \eqref{eq:3.1} to make it independent of Q without shifting VeV at zero temperature~\cite{Quiros:1999jp,Delaunay:2007wb}. Second one is to set Q at a proper scale, like $Q=160(GeV)$ the running value of the top mass, $Q=246.22(GeV)$ EW scale and $Q=1(TeV)$ for supersymmetry purposes. Third one is to take Q as a free parameter to avoid shifting VeV at zero temperature. Here, the last one is considered.

The main idea of the EWPT is that the early universe, which from particle physics point of view may be described by potential \eqref{eq:3.1}, is in a high phase\footnote{In this work, the high (low) phase denotes a phase which is the unstable (stable) vacuum for temperatures below $\mathrm{T_{c}}$.} with \begin{math}VeV=(<h>,<s1>,<s2>)^{high}\end{math} at high temperatures. Cooling down the universe, a new phase appears with \begin{math}VeV=(<h>,<s1>,<s2>)^{low}\end{math}. As the universe cools down, the two phases become degenerate at the critical temperature $T_{c}$. Since the strength of the EWPT is governed by $\xi=\frac{v_{c}}{T_{c}}$, all that needs to be done is to calculate $v_{c}$ and $T_{c}$ from the following conditions:
\begin{equation} \label{eq:3.3}
\begin{split}
& \left.\frac{\partial V_{eff}}{\partial h}\right|_{(<h>,<s1>,<s2>)^{high},T=T_{c}} = 0,\quad \left.\frac{\partial V_{eff}}{\partial h}\right|_{(<h>,<s1>,<s2>)^{low},T=T_{c}} = 0,\\
& \left.\frac{\partial V_{eff}}{\partial s_{1}}\right|_{(<h>,<s1>,<s2>)^{high},T=T_{c}} = 0,\quad \left.\frac{\partial V_{eff}}{\partial s_{1}}\right|_{(<h>,<s1>,<s2>)^{low},T=T_{c}} = 0,\\
& \left.\frac{\partial V_{eff}}{\partial s_{2}}\right|_{(<h>,<s1>,<s2>)^{high},T=T_{c}} = 0,\quad \left.\frac{\partial V_{eff}}{\partial s_{2}}\right|_{(<h>,<s1>,<s2>)^{low},T=T_{c}} = 0,\\
& V_{eff}\Big|_{(<h>,<s1>,<s2>)^{high},T=T_{c}} = V_{eff}\Big|_{(<h>,<s1>,<s2>)^{low},T=T_{c}}.
\end{split}
\end{equation}
The last condition guarantees degeneracy and the others guarantee existence of the high and low vacua. There is no analytical solution for the problem, so the calculations are implemented with the CosmoTransitions package~\cite{Wainwright:2011kj}. The benchmark points and the corresponding results are presented in table \ref{tab:1} and \ref{tab:2}, respectively. Here, the exact calculations are performed by CosmoTransitions to get the effective potential, compared to Ref.~\cite{Tofighi:2015fia} which used the high temperature expansion. Though, the results of Ref.~\cite{Tofighi:2015fia} should be improved for the high temperature expansion case. An extension of the SM with two new scalars was recently studied in Ref.~\cite{Chao:2017vrq}, but there are some differences between it and the present work. First, the high temperature expansion was used in~\cite{Chao:2017vrq}. Second, the cubic term $S_{1}^{3}$, which plays a crucial role in the EWPT as a barrier at tree-level, was not considered in ~\cite{Chao:2017vrq}. %Third, only a two-step phase transition, in which the first-step happens to the dark matter sector, was studied in ~\cite{Chao:2017vrq}.

\begin{table}
  \centering
  \begin{tabular}{| c | c | c | c | c | c | c | c | c | c | c |}
    \hline
        &$M_{1}(GeV)$&$M_{2}(GeV)$&$\theta$&$\chi(GeV)$&$\kappa_{1}$&$\kappa_{3}$&$\kappa_{4}$&$\kappa_{5}$&$\lambda_{2}$&$Q(GeV)$ \\
    \hline
    BM1 &25.27&655.22&-9.80&115&-40.72&0.0528&-4.04&-11.68&0.55&149\\
    BM2 &65.74&337&17.16&65.75&-8.24&0.0241&-34&21.44&0.79&109\\
    BM3 &83.16&235.91&-18.68&69.89&-40.53&0.0132&13.48&-15.44&3.62&160\\
    BM4 &195.89&434.2&-20.42&96.03&-55.38&0.0322&-50.92&47.82&2.36&106.7\\
    BM5 &226.06&126.33&-19.20&54.07&-29.63&0.0016&7.05&5.92&1.79&104.7\\
    BM6 &254.18&420&-15.94&43.82&-35.19&0.0241&-13.01&48.3&1.74&91.19\\
    BM7 &262.86&600&-21.9&53.04&-38.4&0.0618&-2.07&73.55&3.05&91.18\\
    BM8 &305&325&-6&36&-47&0.0012&-2&-26.4&0.13&91.19\\
    \hline
  \end{tabular}
  \caption{Benchmark points which provide the SFOEWPT.}\label{tab:1}
\end{table}

\begin{table}
  \centering
  \begin{tabular}{| c | c | c | c | c | c | c | c | c | c | c |}
    \hline
        &$\mathrm{VeV^{high}_{c}(GeV)}$&$\mathrm{VeV^{low}_{c}(GeV)}$&$\mathrm{T_{c}(GeV)}$&$\xi$ \\
    \hline
    BM1 &(0,6.66,0)&(152.44,58.12,0)&92.61&1.65\\
    BM2 &(0,212.26,0)&(239.05,67.24,0)&60.33&3.96\\
    BM3 &(0,2.13,0)&(117.2,27.16,0)&115.44&1.01\\
    BM4 &(0,191.74,0)&(214.86,100.1,0)&97.06&2.21\\
    BM5 &(0,110.24,0)&(164.43,76.63,0)&114.18&1.44\\
    BM6 &(0,102.36,0)&(215.56,45.91,0)&97.22&2.22\\
    BM7 &(0,113.79,0)&(222.88,48.07,0)&91.84&2.43\\
    BM8 &(0,72.31,0)&(145.52,48.35,0)&118.13&1.23\\
    \hline
  \end{tabular}
  \caption{The values of the VeV of the high and the low phases, $T_{c}$ and the strength of the SFOEWPT.}\label{tab:2}
\end{table}

\section{Gravitational Waves}
\label{sec:4}

The SFOEWPT may justify not only the BAU but also the GW signal produced by the EWPT. Actually, the EWPT occurs at a temperature lower than $T_{c}$, in which the first broken phase bubbles nucleate in the symmetric phase plasma of the early universe. The transition probability is given by $\Gamma(T)=\Gamma_{0}(T)e^{-S(T)}$ where $\Gamma_{0}(T)$ is of order $\mathcal{O}(T^{4})$ and S is the 4-dimensional action of the critical bubbles. For temperatures sufficiently greater than zero, it can be assumed $S=\frac{S_{3}}{T}$ where the 3-dimensional action is given by
\begin{equation} \label{eq:4.1}
S_{3} = 4\pi \int{dr r^{2} \left[ \frac{1}{2}\left(\partial_{r} \vec{\phi}\right)^2 + V_{eff}\right]}.
\end{equation}
Here, $\vec{\phi}=(h,s1,s2)$. The critical bubble profiles, which minimize the action \eqref{eq:4.1}, can be calculated from the equation of motions. The temperature for a particular configuration, which gives the nucleation probability of order $\mathcal{O}(1)$, is the nucleation temperature $T_{n}$.

\begin{figure}
  \centering
  \includegraphics[width=300pt]{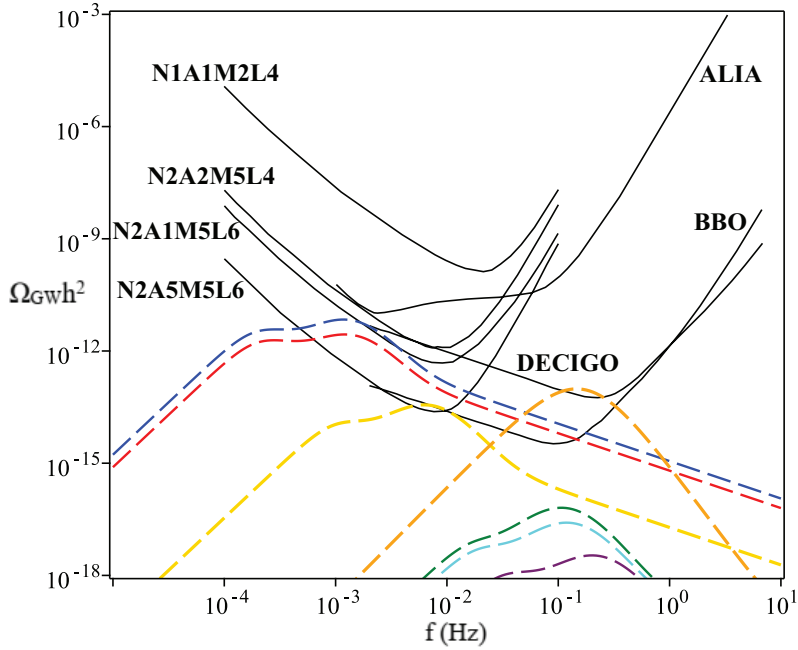}
  \caption{The dashed blue, red, yellow, orange, green, cyan and purple lines represent the GW signal for BM7, BM4, BM6, BM2, BM1, BM5 and BM8, respectively. The solid black lines represent the sensitivity curves and are labeled by the name of the detectors. For eLISA, the sensitivity curves are labeled by the name of the configuration.}\label{fig:1}
\end{figure}

The GW may be produced by the collision of the bubbles at some temperature $T_{*}$, it is usually assumed $T_{*}=T_{n}$. Supposing that the friction force is not enough to prevent the bubbles from running away, the GW signal is given by
\begin{equation} \label{eq:4.2}
\Omega_{GW} h^{2}\simeq\Omega_{col} h^{2}+\Omega_{sw} h^{2}+\Omega_{turb} h^{2}.
\end{equation}
As seen, the GW signal is given by the sum of bubbles collision, sound wave and turbulence in the plasma which respectively read~\cite{Kamionkowski:1993fg,Huber:2008hg,Hindmarsh:2013xza,Hindmarsh:2015qta,Binetruy:2012ze,Caprini:2009yp,Espinosa:2010hh,Caprini:2015zlo}
\begin{equation} \label{eq:4.3}
\begin{split}
& \Omega_{col} h^{2}=1.67 \times 10^{-5} \left(\frac{\beta}{H}\right)^{-2} \frac{0.11\, v_{b}^{3}}{0.42+v_{b}^{2}} \left(\frac{\kappa\, \alpha}{1+\alpha}\right)^{2} \left(\frac{g_{*}}{100}\right)^{-\frac{1}{3}}\frac{3.8\, \left(\frac{f}{f_{col}}\right)^{2.8}}{1+2.8\, \left(\frac{f}{f_{col}}\right)^{3.8}},\\
& \Omega_{sw} h^{2}=2.65 \times 10^{-6} \left(\frac{\beta}{H}\right)^{-1} v_{b} \left(\frac{\kappa_{v}\, \alpha}{1+\alpha}\right)^{2} \left(\frac{g_{*}}{100}\right)^{-\frac{1}{3}} \left(\frac{f}{f_{sw}}\right)^{3} \left(\frac{7}{4+3\left(\frac{f}{f_{sw}}\right)^{2}}\right)^{\frac{7}{2}},\\
& \Omega_{turb} h^{2}=3.35 \times 10^{-4} \left(\frac{\beta}{H}\right)^{-1} v_{b} \left(\frac{\epsilon\, \kappa_{v}\, \alpha}{1+\alpha}\right)^{\frac{3}{2}} \left(\frac{g_{*}}{100}\right)^{-\frac{1}{3}} \frac{\left(\frac{f}{f_{turb}}\right)^{3} \left(1+\frac{f}{f_{turb}}\right)^{-\frac{11}{3}} }{1+\frac{8\pi f}{h_{*}}},
\end{split}
\end{equation}
with
\begin{equation} \label{eq:4.4}
\begin{split}
& f_{col} = 16.5\times10^{-6}\, \mathrm{Hz}\, \left(\frac{0.62}{v_{b}^{2}-0.1\,v_{b}+1.8}\right) \left(\frac{\beta}{H}\right) \left(\frac{T_{n}}{100\, \mathrm{GeV}}\right) \left(\frac{g_{*}}{100}\right)^{\frac{1}{6}},\\
& f_{sw} = 1.9\times10^{-5}\, \mathrm{Hz}\, \left(\frac{1}{v_{b}}\right) \left(\frac{\beta}{H}\right) \left(\frac{T_{n}}{100\, \mathrm{GeV}}\right) \left(\frac{g_{*}}{100}\right)^{\frac{1}{6}},\\
& f_{turb} = 2.7\times10^{-5}\, \mathrm{Hz}\, \left(\frac{1}{v_{b}}\right) \left(\frac{\beta}{H}\right) \left(\frac{T_{n}}{100\, \mathrm{GeV}}\right) \left(\frac{g_{*}}{100}\right)^{\frac{1}{6}},\\
& h_{*} = 16.5\times10^{-6}\, \mathrm{Hz}\, \left(\frac{T_{n}}{100\, \mathrm{GeV}}\right) \left(\frac{g_{*}}{100}\right)^{\frac{1}{6}},\\
& \kappa=1-\frac{\alpha_{\infty}}{\alpha},\\
& \kappa_{v}=\frac{\alpha_{\infty}}{\alpha}\left(\frac{\alpha_{\infty}}{0.73+0.083\sqrt{\alpha_{\infty}}+\alpha_{\infty}}\right),\\
& \alpha_{\infty}=\frac{30}{24\, \pi^2 g_{*}} \left(\frac{v_{n}}{T_{n}}\right)^{2} \left(6\left(\frac{M_{W}}{v}\right)^{2} + 3\left(\frac{M_{Z}}{v}\right)^{2} + 6\left(\frac{M_{top}}{v}\right)^{2} \right).
\end{split}
\end{equation}
$v_{n}$ and $g_{*}$ are the Higgs VeV and the number of the relativistic degrees of freedom at $T_{n}$, respectively. Here, $\epsilon=0.1$, and $g_{*}$ is read from the MicrOMEGAs package~\cite{Belanger:2018mqt,Barducci:2016pcb}. Still, there are three important parameters which should be defined. One of them is the bubble wall velocity, since assumed that the bubbles run away, given by $v_{b} \simeq 1$.
The two others, $\alpha$ and $\beta$, are given as follows
\begin{equation} \label{eq:4.6}
\begin{split}
& \alpha = \left.\frac{\rho_{vac}}{\rho^{*}}\right|_{T_{n}},\\
& \beta = \left.\left[ H\, T\, \frac{d}{dT}\left(\frac{S_{3}}{T}\right) \right]\right|_{T_{n}},
\end{split}
\end{equation}
where $\rho_{vac}=\left( V_{eff}^{high}-T dV_{eff}^{high}/dT \right)-\left( V_{eff}^{low}-T dV_{eff}^{low}/dT \right)$, $\rho^{*}=g_{*}\pi^{2}T_{n}^{4}/30$ and $H_{n}$ are the latent heat (vacuum energy density) released by the EWPT, the background energy density of the plasma and Hubble parameter at $T_{n}$, respectively. Using the CosmoTransitions package~\cite{Wainwright:2011kj}, the parameters $\alpha$, $\beta/H$, $v_{n}$ and $T_{n}$ are calculated and presented in table \ref{tab:3}. In figures \ref{fig:1}, the GW signals are plotted versus frequency for the benchmark points of table \ref{tab:1}. To check if the GW signals for the benchmark points \ref{tab:1} fall within the sensitivity of GW detectors, the sensitivity curves of $\mathbf{eLISA}$, $\mathbf{ALIA}$, $\mathbf{DECIGO}$ and $\mathbf{BBO}$ detectors\footnote{The sensitivity curves of four representative configurations of $\mathbf{eLISA}$ are taken from~\cite{Caprini:2015zlo}. The $\mathbf{ALIA}$, the $\mathbf{DECIGO}$ and the $\mathbf{BBO}$ sensitivity curves are taken from \href{http://gwplotter.com}{GWPLOTTER}. The reader is referred to Ref.~\cite{Moore:2014lga} for details.} are also plotted in the figure \ref{fig:1}. As seen from the figure \ref{fig:1}, the dashed blue line corresponding to the GW signal for the BM7 may be detected by $\mathbf{N2A1M5L6}$ and $\mathbf{N2A5M5L6}$ configurations of $\mathbf{eLISA}$ and $\mathbf{BBO}$ detectors. The dashed red and yellow lines corresponding, respectively, to the GW signal for the BM4 and the BM6 may be detected by $\mathbf{N2A5M5L6}$ configuration of $\mathbf{eLISA}$ and $\mathbf{BBO}$ detectors. The dashed orange line corresponding to the BM2 may be detected by $\mathbf{DECIGO}$ and $\mathbf{BBO}$ detectors. The dashed green, cyan and purple lines corresponding, respectively, to the GW signal for BM1, BM5 and BM8 cannot be detected by the mentioned detectors. The GW signal for the BM3 isn't big enough to be shown at the scale of the figure \ref{fig:1}.

\begin{table}
  \centering
  \begin{tabular}{| c | c | c | c | c | c | c | c | c | c | c |}
    \hline
        &$\mathrm{VeV^{high}_{n}(GeV)}$&$\mathrm{VeV^{low}_{n}(GeV)}$&$\mathrm{T_{n}(GeV)}$&$\alpha$&$\beta/H$ \\
    \hline
    BM1 &(0,6.46,0)&(169.49,68.13,0)&89.15&0.0324&6291.32\\
    BM2 &(0,212.75,0)&(244.78,63.20,0)&41.61&0.2595&18459.13\\
    BM3 &(0,2.09,0)&(127.05,30.59,0)&114.43&0.0119&27039.55\\
    BM4 &(0,194.67,0)&(243.13,93.35,0)&51.83&0.3131&130.43\\
    BM5 &(0,110.54,0)&(185.04,69.54,0)&110.97&0.0245&5644.92\\
    BM6 &(0,103.24,0)&(234.55,41.18,0)&77.05&0.0890&433.45\\
    BM7 &(0,114.91,0)&(243.12,42.47,0)&43&0.5388&150.54\\
    BM8 &(0,72.67,0)&(157.93,45.79,0)&115.38&0.0175&9306.63\\
    \hline
  \end{tabular}
  \caption{The values of the VeV of the high and the low phases, $T_{n}$, $\alpha$ and $\beta/H$.}\label{tab:3}
\end{table}

\begin{figure}[!h]
     \subfloat[\label{subfig-1:dummy}]{%
       \includegraphics[width=0.5\textwidth]{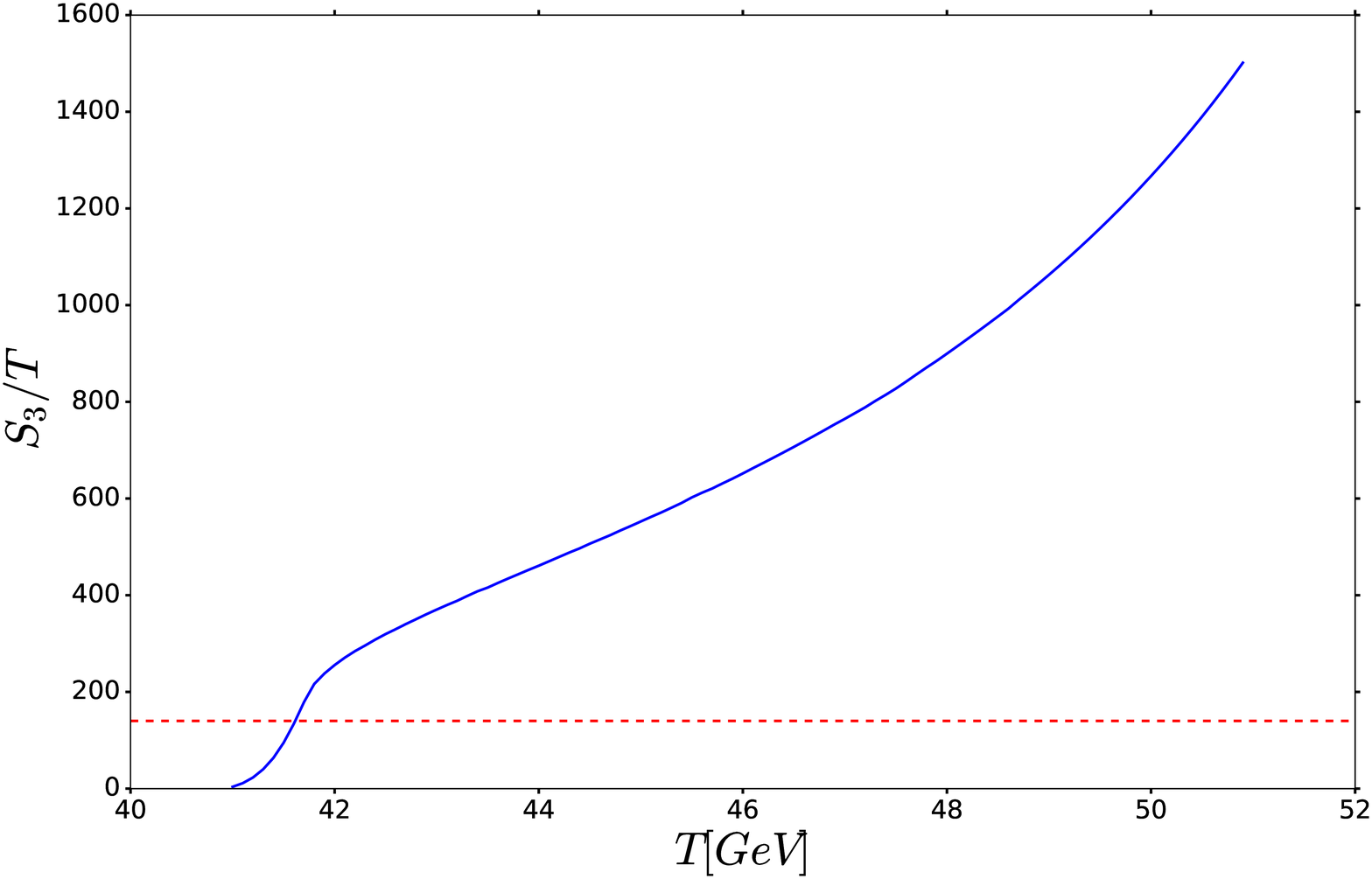}
     }
     \hfill
     \subfloat[\label{subfig-2:dummy}]{%
       \includegraphics[width=0.5\textwidth]{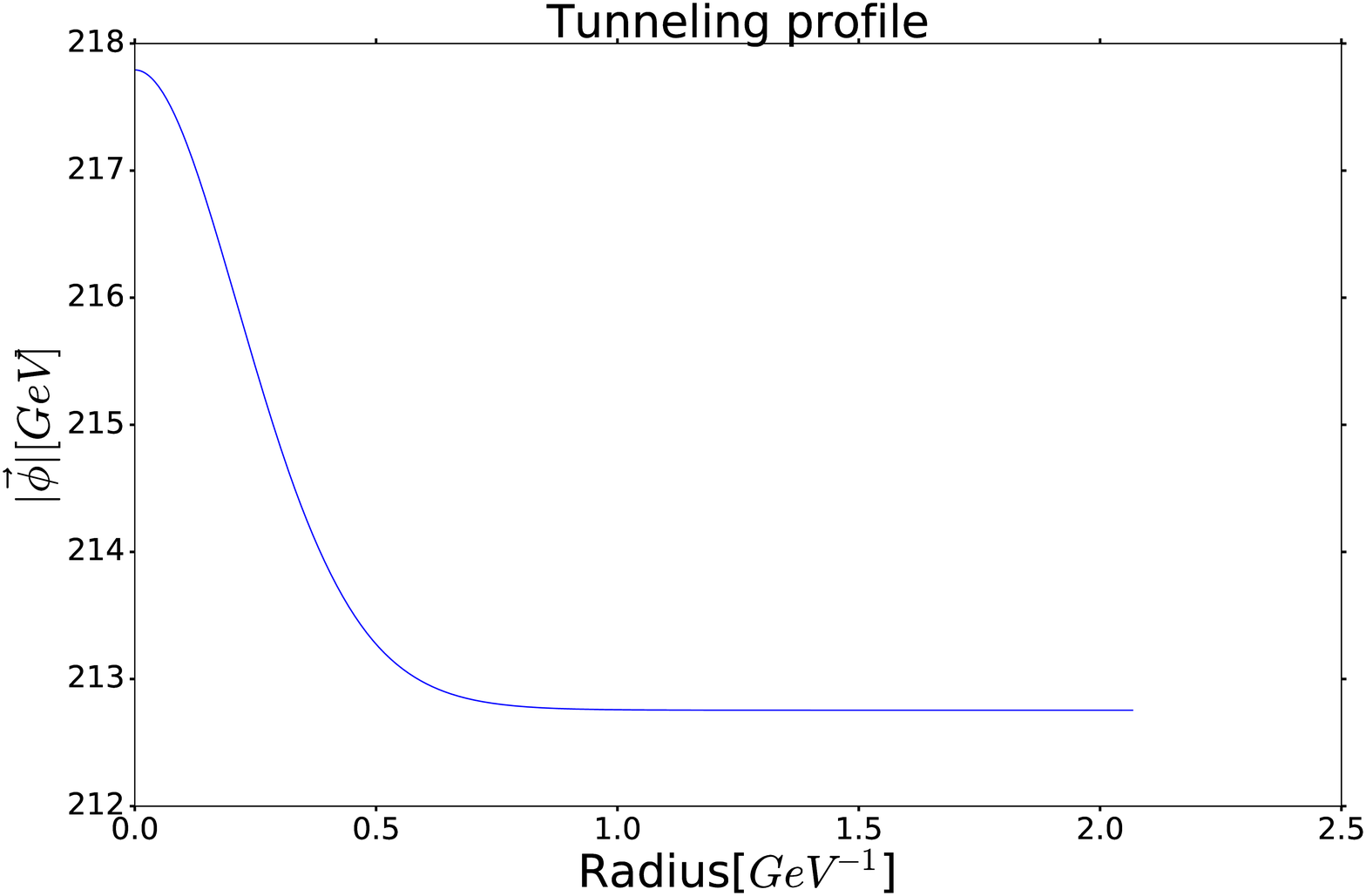}
     }
     \hfill
     \subfloat[\label{subfig-3:dummy}]{%
       \includegraphics[width=0.5\textwidth]{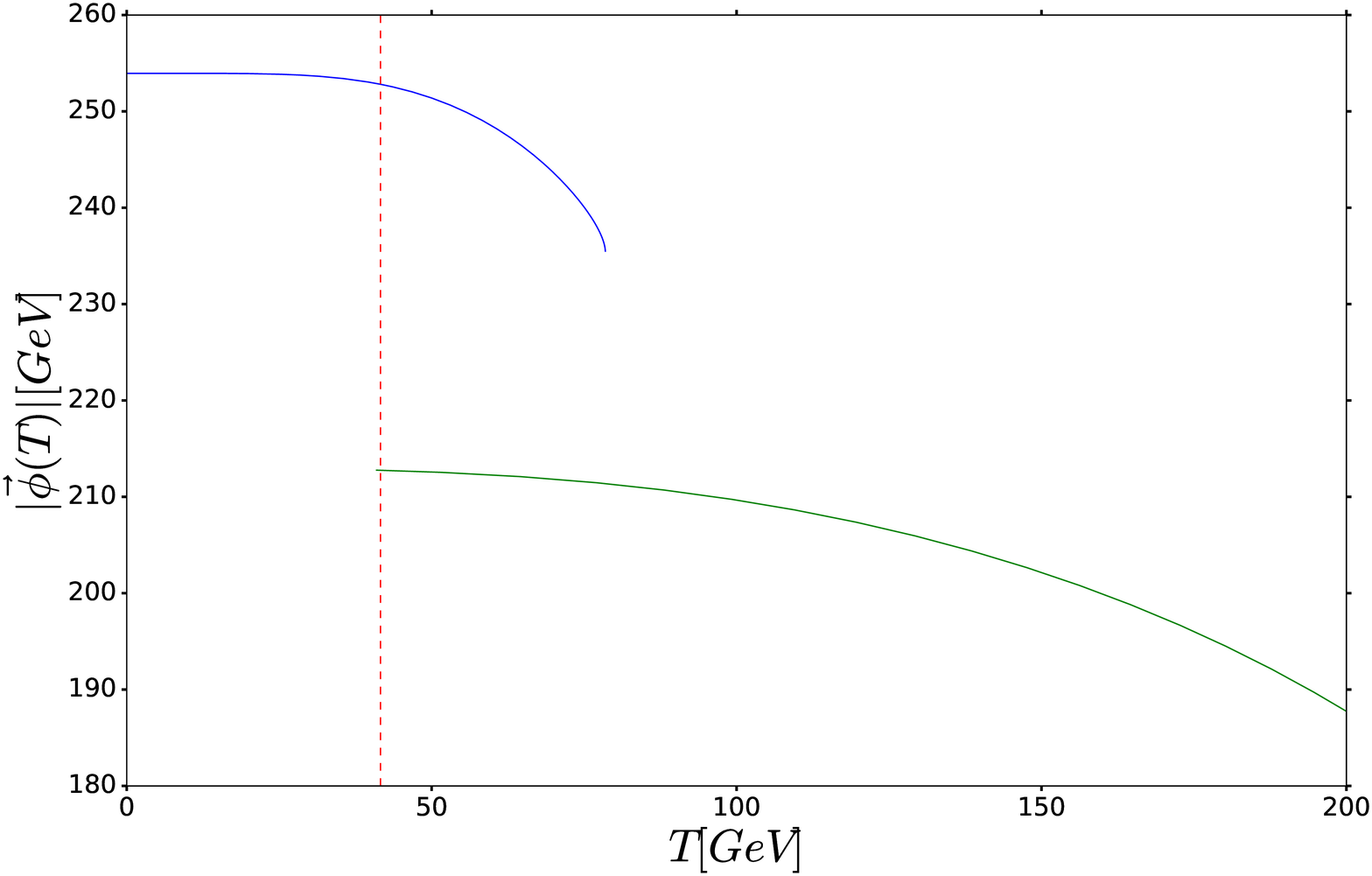}
     }
     \hfill
     \subfloat[\label{subfig-4:dummy}]{%
       \includegraphics[width=0.5\textwidth]{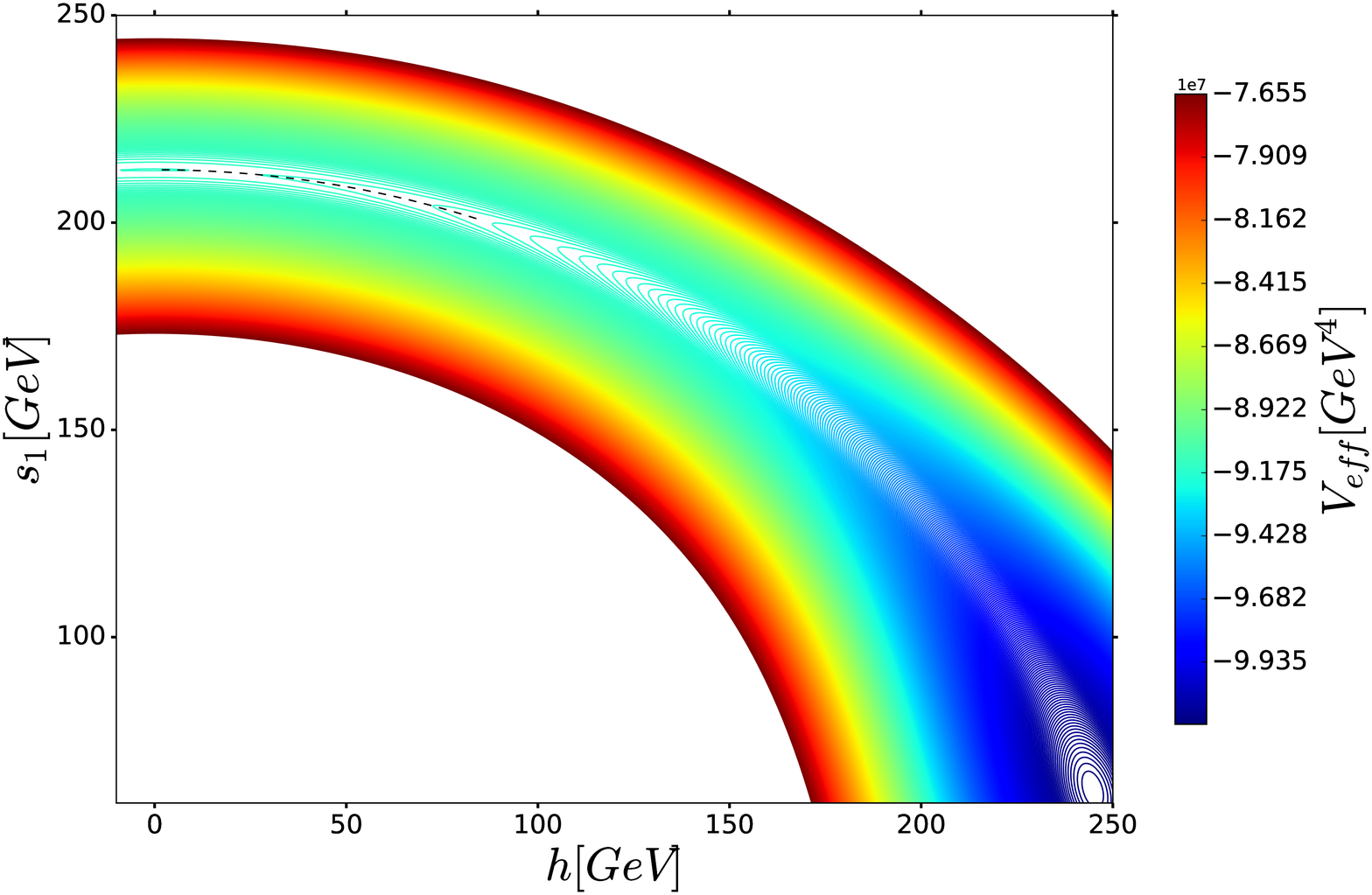}
     }
     \caption{Phase transition properties of BM2: The subfigure (a) presents $S_{3}/T$ versus temperature, the dashed horizontal red line shows $S_{3}/T=140$ where nucleation occurs. The subfigure (b) presents the tunneling profile as a function of radius. The subfigure (c) presents the norms of high (green line) and low (blue line) phases as functions of temperature, the dashed vertical red line shows the nucleation temperature. The subfigure (d) presents the contour levels of the potential at the nucleation temperature $T_{n}=41.61 (GeV)$, the dashed black line shows the tunneling path.}
     \label{fig:2}
   \end{figure}

According to the tables \ref{tab:2} and \ref{tab:3}, it seems that BM2 is a special point. The value of $\beta/H$ is large at this point, while, the nucleation temperature is not very close to the critical temperature. At the same time, $T_{n}$ is low and $\alpha$ is large.\footnote{The authors thank an anonymous referee for pointing this out.} To clarify the situation of BM2, the phase transition properties of BM2 are shown in the figure \ref{fig:2}. As seen from the subfigure \ref{fig:2}-(a), the slope of $S_{3}/T$ increases around $T_{n}$ which indicates the parameter $\beta/H$ is large. The physics of this situation can be described by the tunneling profile, the norm of phases as a function of temperature, and the contour levels of the potential with the tunneling path. As seen from subfigure \ref{fig:2}-(b) and \ref{fig:2}-(d), the center of bubble is far away from the stable vacuum. Also from subfigure \ref{fig:2}-(c), it is seen that the transition occurs at a temperature where the unstable vacuum is close to disappearance. The values of potential at high and low phases are, respectively, $V^{high}_{eff}=-91583128.19 (GeV^{4})$ and $V^{low}_{eff}=-101840540.47 (GeV^{4})$, which give the pressure difference $\Delta p = 10257412.28 (GeV^{4})$. The barrier location is at $(h, s_{1}, s_{2})=(19.84, 212.19, 0)$ with $V_{eff}=-91582001.71 (GeV^{4})$ which gives the barrier height $\Delta V_{barrier \: height}=1126.48 (GeV^{4})$. Clearly, the barrier height is very small, $\Delta V_{barrier \: height}/\Delta p = 0.0001$. Due to the reasons given above, the bubbles are extremely thick walled. Since the barrier height is very small, the transition duration is very short, accordingly, the parameter $\beta/H$ is quite large. This extremely thick walled case is similar to the second-order phase transition in which $\beta/H \rightarrow \infty$ and there is no barrier. Moreover, there is another interesting note for BM2. Due to the cubic term $s_{1}^{3}$, it is expected that the model has a sizable barrier at tree-level like the supercooled scenario discussed in~\cite{Kobakhidze:2017mru}, but this is not the case for BM2. At this point, the model mimics the behavior of
supercooled phase transitions with the supercooling parameter $(T_{c}-T_{n})/T_{c}=0.31$, though, the transition is short-lived.

\section{Dark Matter}
\label{sec:5}

As mentioned prior, imposing the $Z_{2}$ symmetry on $S_{2}$ makes it a viable candidate for the DM. Considering the freeze-out formalism, the DM relic density abundance can be calculated by solving the Boltzmann equation,
\begin{equation} \label{eq:5.1}
\frac{dn}{dt}=-3 H n-<\sigma v> (n^{2}-n_{eq}^{2}),
\end{equation}
 where n, H, \text{<$\sigma$v>} are the number of the DM particles, the Hubble parameter and the thermally-averaged cross section for the DM annihilation, respectively. It is customary to rewrite the Boltzmann equation in terms of $Y=n/s$, where $s$ is the total entropy density of the universe, the result is~\cite{Gondolo:1990dk}
\begin{equation} \label{eq:5.2}
\frac{dY}{dx}=-\left(\frac{45\,G\,g_{*}}{\pi}\right)^{-\frac{1}{2}}\frac{M\,h_{*}}{x^{2}}(1+\frac{1}{3}\frac{T}{h_{*}}\frac{dh_{*}}{dT})<\sigma v>(Y^{2}-Y_{eq}^{2}),
\end{equation}
where $x=M/T$, $M$ is the DM mass. $h_{*}$ is the effective degree of freedom for the entropy densities. The DM relic density abundance reads,
\begin{equation} \label{eq:5.3}
\Omega_{DM} h^{2}\simeq (2.79\pm 0.05) \times 10^{8} \left(\frac{M}{GeV}\right) Y(0).
\end{equation}
It is assumed the usual SM particles only interact with Higgs in this model, so the annihilation channels for DM via the Higgs portal s-channel are $s_{2}s_{2}\rightarrow W^{+}W^{-},ZZ,f\bar{f}$. Also, there exists $s_{2}s_{2}\rightarrow \phi_{i}\phi_{j}$ (with $\phi_{i(j)}=h,s_{1}$ and $i(j)=1,2$) via $s$, $t$ and $u$ channels and four-point interactions.

 The parameter space is constrained by the direct detection DM searches. To do this, one needs to calculate the spin-independent cross section for DM-nucleon scattering\footnote{The spin-dependent case is not studied here, because the DM candidate is assumed to be a scalar.}, and compares the result with the $\mathbf{XENON1T}$ $\mathbf{2018}$ experiment data~\cite{Aprile:2018dbl}. The spin-independent cross section is given by
\begin{equation} \label{eq:5.4}
\sigma_{SI}=\frac{4 M_{s_{2}}^{2} M_{N}^{2}}{\pi (M_{s{2}}+M_{N})^{2}}\Big| \mathcal{M}_{s_{2}-N} \Big|^{2},
\end{equation}
where $M_{s_{2}}$, $M_{N}$ and $\mathcal{M}_{s_{2}-N}$ are the DM mass, the nucleus mass and the scattering amplitude at low energy limit, respectively. $\mathcal{M}_{s_{2}-N}$ is related to $\mathcal{M}_{s_{2}-quark}$, so, calculating effective Lagrangian coefficients and nucleon form factors, $\mathcal{M}_{s_{2}-N}$ can be obtained from $\mathcal{M}_{s_{2}-quark}$. Here, the model is implemented in SARAH~\cite{Staub:2008uz,Staub:2013tta,Staub:2009bi}, the model spectrum is obtained by SPheno~\cite{Porod:2011nf,Porod:2003um} and the DM properties are studied by MicrOMEGAs~\cite{Belanger:2018mqt,Barducci:2016pcb}. The results are presented in table \ref{tab:4}.
\begin{table}
  \centering
  \begin{tabular}{| c | c | c | c | c | c | c | c | c | c | c |}
    \hline
        &$\Omega_{DM} h^{2}$&$\sigma_{SI}^{proton}(pb)$&$\sigma_{SI}^{neutron}(pb)$\\
    \hline
    BM1 &$0.104$&$5.151\times10^{-12}$&$5.315\times10^{-12}$\\
    BM2 &$0.12$&$6.469\times10^{-12}$&$6.667\times10^{-12}$\\
    BM3 &$6.28\times10^{-2}$&$9.447\times10^{-12}$&$9.729\times10^{-12}$\\
    BM4 &$0.109$&$5.038\times10^{-11}$&$5.195\times10^{-11}$\\
    BM5 &$0.108$&$6.186\times10^{-12}$&$6.363\times10^{-12}$\\
    BM6 &$0.12$&$4.223\times10^{-12}$&$4.354\times10^{-12}$\\
    BM7 &$5.90\times10^{-2}$&$4.755\times10^{-11}$&$4.906\times10^{-11}$\\
    BM8 &$0.12$&$2.480\times10^{-11}$&$2.556\times10^{-11}$\\
    \hline
  \end{tabular}
  \caption{The values of the DM relic abundance and the spin-independent cross sections.}\label{tab:4}
\end{table}
As seen in table \ref{tab:4}, the relic density of all benchmark points is compatible with $\mathbf{Planck}$ $\mathbf{2018}$ data which reports $\Omega_{c}h^{2}=0.120\pm0.001$\footnote{$0.05\leq \Omega_{c}h^{2}\leq 0.12$ would be acceptable.}. Moreover, the results fit with $\mathbf{XENON1T}$ $\mathbf{2018}$ experiment which gives an upper limit, less than $\mathbf{LUX}$ $\mathbf{2017}$~\cite{Akerib:2016vxi} and $\mathbf{PandaX}$-$\mathbf{II}$ $\mathbf{2017}$~\cite{Cui:2017nnn} reports, on the DM-nucleon spin-independent elastic scattering cross section. In the DM study, there are two differences with Ref.~\cite{Chao:2017vrq}. The first is the $s_{1}^{3}$ interaction which gives a significant contribution to the DM annihilation through $s_{1}$ s-channel, and consequently to relic density. The second is the spin-independent cross section which was taken to be zero in Ref.~\cite{Chao:2017vrq}, but the more realistic case like here is to have a non-zero DM-nucleon cross section, if weakly interacting massive particles (WIMP) constitute the DM. This is the main idea behind $\mathbf{LUX}$, $\mathbf{PandaX}$-$\mathbf{II}$ and $\mathbf{XENON1T}$ experiments.

\section{Conclusions}
\label{sec:6}

The main goal of this work has been to investigate the EWPT, the GW and the DM issues in an extension of the SM by adding two scalar degrees of freedom. To reach the goal, it has been assumed that one of the new scalars has a non-zero VeV to assist the phase transition and the other has no VeV to be a viable DM candidate. It has been seen if one takes the most general renormalizable form of the potential, the model can represent all the signals together. As seen from tables \ref{tab:2} and \ref{tab:3}, the model can have phase transitions from strong ($\xi \sim 1$) to very strong ($\xi \sim 4$). From figure \ref{fig:1}, the model presents the GW signals from the frequency range of $10^{-5}$(Hz) to $10$(Hz) which are detectable by $\mathbf{eLISA}$, $\mathbf{BBO}$ and $\mathbf{DECIGO}$. From table \ref{tab:4}, the model provides the DM signals which are in agreement with the $\mathbf{Planck}$ $\mathbf{2018}$ data and the $\mathbf{XENON1T}$ $\mathbf{2018}$ experiment. It is seen that the DM candidate may be quite massive with a mass greater than \text{100(GeV)} which belongs to the extremely cold DM; although, since the model has a rich parameter space, the lighter DMs might be found by performing a Monte Carlo simulation via a computer cluster. With all of these, it can be concluded that the SFOEWPT, the GW and the DM signals can successfully be described by the present model as an extension of the SM with two additional real gauge singlet scalars. As a final note, it has been assumed that the GW production from bubble collisions follows from thin-wall and envelope approximations which is usual in the literature. In this assumption, only uncollided parts of the bubbles are taken into account as the GW sources. Recently, it has been shown that the GW production from bubble collisions is analytically solvable~\cite{Jinno:2016vai,Jinno:2017fby}. Also, the possibility of using GWs and collider experiments to constrain the EWPT has been discussed in~\cite{Hashino:2018wee}. It is left for future work to study the GW signals of the present model using these recent studies.

\section{Acknowledgements}
\label{sec:7}
The authors would like to thank Ryusuke Jinno for comments on the GW production from bubble collisions.


\bibliographystyle{JHEP.bst}
\bibliography{paper201902}

% Please avoid comments such as "For a review'', "For some examples",
% "and references therein" or move them in the text. In general,
% please leave only references in the bibliography and move all
% accessory text in footnotes.

% Also, please have only one work for each \bibitem.
\end{document}